\documentstyle[epsf,prd,aps]{revtex}
\begin{document}
\draft
\title
{ The Jang equation, apparent horizons, and the Penrose inequality }
\author
{ Edward Malec$^{*1}$ and Niall \'O Murchadha$^{+1}$}
\address{ $^*$ Institute of Physics,  Jagiellonian University,
30-059  Cracow, Reymonta 4, Poland.}
\address{$^{+}$ Physics Department, University College, Cork, Ireland.}
\address{$^1$ Erwin Schr\"odinger International Institute, Boltzmanngasse 9, A-1090
Vienna, Austria.}
\maketitle
\begin{abstract}
The Jang equation in the spherically symmetric case reduces to a first order equation.
This permits an easy analysis of the role apparent horizons play in the (non)existence
of solutions. We demonstrate that the proposed derivation of the Penrose inequality
based on the Jang equation cannot    work in the spherically symmetric case. Thus it is
fruitless to apply this method, as it stands, to the general case.
We show also that those analytic criteria for the formation of horizons that are based
on the use of the Jang equation are of limited validity for the proof of the trapped
surface conjecture.
\end{abstract}

\section{Introduction}

P. S. Jang introduced his eponymous equation in 1978
in the context of the initial value formulation of
general relativity \cite{Jang}.  The initial data for the Einstein equations
consists of a quartet of objects $(g_{ij}, K^{ij}, \rho, J^i)$
which satisfy the constraints. Many of the difficult issues in this area
are easier to express and solve if one knows that the scalar curvature
of the metric $g_{ij}$ is positive. Unfortunately, one has no guarantee
that this is valid for a given initial data set.

 The Jang equation is a
non-linear second order elliptic equation for a scalar $f$
which depends on $g_{ij}$ and $K^{ij}$. Given a metric $g_{ij}$, and a solution $f$, one
constructs a related metric ${\bar g}_{ij} = g_{ij} + f_{,i}f_{,j}$. Jang showed that
the scalar curvature of the new metric has nice positivity properties.

The first precise results about the Jang equation were derived by Schoen and Yau
in their second article on the positivity of the total energy of the gravitational
field \cite{SY2}. In this article they proved that if asymptotically flat initial data
do not contain any apparent horizon (either future or past) a
regular solution must exist. The converse result, i.e., if a
regular solution to the Jang equation does not exist, then the data must contain
apparent horizons, played a key role in their article `Existence of a black hole due to
condensation of matter'\cite{SY3}. Given a
Riemannian metric with a nonzero scalar curvature $^{(3)}R$, one can find a
conformally related manifold with zero scalar curvature by solving the Lichnerowicz
equation, $8\nabla^2 \phi - ^{(3)}R\phi = 0$. Schoen and Yau also showed (in \cite{SY2})
 that one could always solve the Lichnerowicz equation on the `Jang transformed' metric
because of the positivity property discovered by Jang.

The Penrose inequality relates the area of the outermost apparent horizon to the ADM
mass. It has been suggested that the Jang equation, in a method described in Sec. IV,
can be used  to prove the Penrose inequality. We explicitly demonstrate, analysing spherically symmetric initial data, that one cannot control simultaneously the
mass of the three-manifold  and the area in the various steps of this
construction. As a result, the Penrose inequality cannot be derived by this method even
in the spherically symmetric case, and therefore it could not be effective in the
general case. It remains an open question whether a radical alteration of the method
would give the desired result.

In the next sections we  focus our attention on the Jang
equation with   spherically symmetric data  and  seek a
spherically symmetric solution. In this case it reduces to a first
order equation (essentially for the radial derivative of $f$) and it is much easier
to determine whether solutions do or do not exist. In the sections following Sec. IV we
investigate the validity of the Jang-equation based methods for diagnosing the presence
of trapped surfaces. The net conclusion we draw from this analysis is that the Jang
equation is not very useful as either a predictor or finder of apparent horizons, and
that it is not particularly suitable for the proof of the trapped surface conjecture.

\section{Spherically symmetric Jang equation}

As described by Schoen and Yau, one starts with a pair $(g_{ij}, K^{ij})$. One extends
the three metric to a four metric (which is Riemannian, not pseudoriemannian!) by adding
trivially a fourth coordinate (call it $w$) such that $g_{ww} = +1, g_{wi} = 0$, i.e.,
  \[  g_{\mu \nu}= \left(
 \matrix{ 1, & 0  \cr 0, & g_{ij}} \right) \]

 In this four manifold
we find a three surface defined by $w = f(x^i)$. The Jang equation is that the
mean curvature of this three surface equals the trace of $K_{ij}$, taken with respect
to the induced three metric. This gives a second-order, nonlinear elliptic equation for
the scalar $f$.

Let us assume that both the three metric and the extrinsic curvature are spherically
symmetric. This means that we can write the metric as
\[
ds^2 = g_{rr}dr^2 + R^2(r)d\Omega^2
\]
where $r$ is some radial coordinate and $R$ is the areal (Schwarzschild) radius of
the isometry spheres. The spherically symmetric extrinsic curvature can be written as
\[
K^{ab} = n^an^bK_l + (g^{ab} - n^an^b)K_{R}
\]
where $K_l$ and $K_R$ are two scalars and $n^a$ is the outward pointing unit normal to
the surfaces of constant $R$.

We make the following coordinate transformation:
 \begin{eqnarray*}
 \bar w &= w - f(r)\\ 
\bar r &= r\\ 
\bar \theta &= \theta \\ 
\bar \phi &= \phi
 \end{eqnarray*}
 where the $\bar w = 0$ surface is the slice we are
 interested in. The transformed metric becomes
 \[ \bar g_{\mu \nu}= \left(
 \matrix{ 1, & f', & 0, & 0 \cr f', & g_{rr} + f'^2, & 0, & 0\cr 0 & 0
 & R^2 & 0 \cr 0 & 0 & 0 & R^2\sin^2\theta \cr} \right) \]
 and
\[ \bar g^{\mu
 \nu}= \left( \matrix{ 1+ f'^2g^{rr}, & -f'g^{rr},  & 0 & 0\cr
 -f'g^{rr}, & g^{rr}, & 0 & 0
 \cr 0 & 0 & {1 \over R^2} & 0 \cr 0 & 0 & 0 & {1 \over R^2\sin^2\theta}
 \cr} \right) \]
 where the prime denotes derivative with respect to $r$.

The unit normal to the surface defined by $\bar{w} = w - f = 0$ is given by
\[
(\bar{n}^{\mu}) = {\Bigl( 1 + f'^2g^{rr}, -f'g^{rr}, 0, 0 \Bigr) \over \sqrt{1 +
f'^2g^{rr}}},
\]
and the mean curvature of the slice is given by
\begin{eqnarray}
H &= \bar{n}^{\mu}_{;\mu} = {1 \over \sqrt{^{(4)}\bar{g}}}(
\sqrt{^{(4)}\bar{g}}\bar{n}^\mu)_{,\mu}
\cr &= -{\sqrt{g^{rr}} \over R^2}\left({R^2 f'\sqrt{g^{rr}} \over \sqrt{1 + f'^2
g^{rr}}}\right)_{,r}. 
\end{eqnarray}

The induced three metric is given by
\[ \bar g^{ab}= \left( \matrix{ {1 \over g_{rr} + f'^2}, &  0 & 0\cr
  0 & {1 \over R^2} & 0 \cr  0 & 0 & {1 \over R^2\sin^2\theta}
 \cr} \right) \]
 and the relevant trace of $K$ is given by
 \begin{eqnarray}
 \bar{g}^{ab}K_{ab} &= {K_l \over  1 + f'^2g^{rr}} + 2K_R \cr
 &= (K_l + 2K_R) - K_l { f'^2g^{rr} \over  1 + f'^2g^{rr}}\cr
 &= { \rm tr}K - K_l { f'^2g^{rr} \over  1 + f'^2g^{rr}}.
 \end{eqnarray}

 The spherically symmetric Jang equation is
 \begin{equation}
 {\sqrt{g^{rr}} \over R^2}\left({R^2 f'\sqrt{g^{rr}} \over \sqrt{1 + f'^2 g^{rr}}}\right)_{,r}
+
  { \rm tr}K - K_l { f'^2g^{rr} \over  1 + f'^2g^{rr}} = 0.
\label{main}
  \end{equation}

  This equation can be simplified by introducing
  \begin{equation}
  k =  {f'\sqrt{g^{rr}} \over \sqrt{1 + f'^2 g^{rr}}} = {f_{,l} \over \sqrt{1 +
f_{,l}^2}} \Rightarrow f_{,l} = {k \over \sqrt{1 - k^2}}, \label{k}
  \end{equation}
  where $l$ is the unit proper distance in the radial direction, as a new variable. The
Jang
  equation now  becomes
  \begin{equation}
  (R^2k)_{,l} + R^2[{\rm tr}K - K_lk^2] = 0.  \label{J}
  \end{equation}

  The apparent horizon conditions are
  \begin{equation}
  R_{,l} \pm RK_R = 0.\label{ah}
  \end{equation}
with the plus sign for a future apparent horizon, and the minus for a past.
  
  It is not immediately clear what  role apparent horizons play in the disappearance of
regular
  solutions to Eq.\ (\ref{J}). What is obvious, however, is that the Jang equation does
not
  distinguish between future and past horizons. If one reverses the sign of the extrinsic
  curvature in the initial data, one will get the same solution, $k$, for Eq.\ (\ref{J})
but with the  sign reversed. It is clear from Eq.\ (\ref{k})
 that $f_{,l} \rightarrow \pm \infty$ as
$k  \rightarrow \pm 1$. Therefore the blowup occurs when the
derivative
  of the solution goes to infinity and this is insensitive to the sign of the extrinsic
curvature.    On the other hand, if $k$ rises above $+1$, then from Eq.
(\ref{J3}) one can see that there  must be future trapped surfaces.
Similarly, if $k$ decreases below $-1$, then there exist past trapped
surfaces.

  The great advantage of this analysis is that we have reduced the Jang equation to a fairly
  simple first order equation and that one  can now identify quite easily when the
solution becomes unphysical. We confirm the Schoen-Yau result that blowup can only occur
when the initial data have  apparent horizons.
In Sections V and VI we will apply this equation in a number of simple situations.  It
will be shown, by considering special families of initial data,
  that blowup can occur far beyond the point where the first apparent horizon appears.
   We also construct families where, despite the existence of apparent horizons, no
blowup ever occurs.

\section{Uniqueness and apparent horizons}

We can rewrite  Eq.\ (\ref{J}) in the following way:
\begin{equation}
R^2k_{,l} + 2R[R_{,l}k + RK_R] + R^2K_l(1 - k^2) = 0.
\label{J1}
\end{equation}
Let us assume that there exists a solution to this equation for data without either
a future or past apparent horizon. Further assume that at some point $|k| <
1$. For example, at a regular center one needs that $k = 0$ at $R = 0$ (otherwise,
since $k$ is the derivative of
$f$, there would be a conical singularity at the origin). We can show, by method of
contradiction, that
$ -1< k < +1$.  We assume that
$k$ starts off less than 1 and rises up through +1. At $k = 1$ Eq.\ (\ref{J1}) becomes
\begin{equation}
R^2k_{,l} + 2R[R_{,l} + RK_R]  = 0. \label{J2}
\end{equation}
Since no future apparent horizons exist, we have $R_{,l} + RK_R > 0$. Therefore, from
Eq.\ (\ref{J2}), one gets $k_{,l} < 0$ which contradicts the fact that it must rise up
through $k = 1$. 

Alternatively, one could have $k$ dropping down through $-1$. At $k = -1$ Eq.\ (\ref{J1})
becomes
\begin{equation}
R^2k_{,l} + 2R[-R_{,l} + RK_R]  = 0. \label{J3}
\end{equation}
Since we assume no past apparent horizons, $R_{,l} - RK_R > 0$. Therefore from
Eq.\ (\ref{J3}), follows $k_{,l} > 0$ which contradicts the fact that it must drop down
through $k = -1$. Hence, if a solution exists, it must lie between $-1$ and +1. In turn,
this means that $f_{,l}$ is well-defined and it can be integrated to find $f$ itself.

Let us consider a `critical' initial data set on the boundary between `good' and `bad'
solutions to the Jang equation.  This will be a set where the maximum value of $k$
equals  +1 (or the minimum equals
$-1$). Therefore we have a point in this critical data set where $k =
\pm 1$ and $k_{,l} = 0$. If this is substituted into Eq.\ (\ref{J1}) one gets that the
point, where the equation breaks down, must satisfy
\begin{equation}
\pm R_{,l} + RK_R = 0,
\end{equation}
i.e., at an apparent horizon. The reader should be warned that this is a very special
case. While the Jang equation ceases to have regular solutions only when there are
apparent horizons, the points where the gradient of $f$ goes infinite will almost always
not coincide with apparent horizons.

Let us assume that a regular solution exists.
We can show that this solution is unique.
Assume that the equation (\ref{J}) possesses two regular solutions $k_1$ and $k_2$
that coincide at an initial point $l_0$ (possibly $l_0=0$).
Define $Y\equiv R^2(k_1-k_2)$; one easily finds that 
\begin{equation}
Y_{,l}=-(2K_R-trK)(k_1+k_2) Y.
\label{unique}
\end{equation}
One
has, after a little calculation and a Gronwall  type argument, that $|Y(l)|\le
|Y(l_0)|
\exp \Bigl( \int_{l_0}^ldl |(2K_R-trK)(k_1+k_2)|\Bigr)$. We know that $(k_1+k_2)$ is
finite and let us assume that the extrinsic curvature within the integrand is regular
enough to ensure that the integral converges.
Since $Y(l_0)=0$, one immediately sees that $Y$ is identically zero and $k_1=k_2$.
Note that this uniqueness result holds without any assumption about trapped surfaces.

\section{The Jang equation and the Penrose inequality}

Jang quite rightly pointed out that if his equation was sensitive
to the existence of apparent horizons (he specifically expected that regular  solutions
could be  absent in such a case), then it may be useful in proving the Penrose
inequality \cite{Jang}. The Penrose inequality is a statement relating the asymptotic
mass of a Cauchy hypersurface and the apparent horizon area on this hypersurface.  It has
been proven only in special cases -- in the ``Riemannian case" (\cite{HI},
\cite{Bray}) and in spherically symmetric spacetimes \cite{MOM94}. The   existing proofs
(or  scenarios for the proofs \cite{hay}, \cite{Frau} and \cite{MMS}; see also references
therein) are heavily based on the fact that the Hawking mass of an apparent horizon
itself satisfies the Penrose inequality and that is monotonic in a very special class of
foliations (the inverse mean curvature foliations).

It has been widely suggested that a three-step approach might allow one to prove the
Penrose inequality in the general case. Let us assume that one is given an
asymptotically flat initial data set with an outermost apparent horizon. One first
solves the Jang equation in the region between the outermost horizon and infinity. Since
this region has no horizons, there must be a regular solution to the Jang equation.
The second step is to perform the Jang transformation and obtain a metric whose scalar
curvature has a positivity property -- assuming in addition the dominant energy condition
for material fields -- that guarantees that the
Lichnerowicz equation has a solution (see \cite{SY2}).
 This allows us to conformally transform to a manifold with
  zero scalar curvature. Then - provided that the original
apparent horizon transforms into an outermost minimal surface - 
one can be able to use the inverse mean curvature flow argument of Huisken and Ilmanen
 \cite{HI} and show that the Penrose inequality
holds. We argue that this procedure cannot be implemented even in the spherical case and
therefore there is no hope of doing so in the general, nonspherical, case.

Let us have a spherical set of initial data with suitable decay at infinity and identify
the outermost apparent horizon. For concreteness, let us assume it is a future apparent
horizon. Therefore on this surface we have
 \begin{equation}
  R_{,l} + RK_R = 0,\hskip 1cm R_{,l} - RK_R \ge 0.\label{ah1}
  \end{equation}
Adding the two conditions, we immediately get $R_{,l} \ge 0$. This means that the mean
curvature of the outermost apparent horizon is nonnegative. The inverse
mean curvature flow argument of Huisken and Ilmanen \cite{HI} requires that the
starting surface have zero mean curvature. Thus this condition has to be
maintained under whatever transformations are made to the manifold. We end up comparing a
final inner area to a final asymptotic mass. Since one really wants to compare the
initial area to the initial mass, we do not want to make any changes which either reduce
the inner area or increase the mass. The mean curvature of every spherical surface
outside the outermost horizon is positive in spherically symmetric geometries. A minimal
surface with $K_R = 0$ is an apparent horizon, while a minimal surface with $K_R \ne 0$
is either past or future trapped so must have an apparent horizon outside it.

We first need to solve the Jang equation, Eq.\ (\ref{J1}), i.e.,
\begin{equation}
R^2k_{,l} + 2R[R_{,l}k + RK_R] + R^2K_l(1 - k^2) = 0.
\end{equation}
between the horizon and infinity. The only uncertainty is in the choice of boundary
conditions at the horizon. These will be dictated by the  demand that the method of 
Huisken and  Ilmanen works; and that works in turn only if the outermost horizon actually
is a minimal surface. It happens that only $k=1$ and $k=-1$ do the trick; starting from
a three-manifold with an apparent horizon in the first step, one ends with a three-manifold
having  a minimal surface in  the third step.

Let us first see what happens when we pick $k = 1$. At the horizon  $R_{,l} +
RK_R = 0$ so that $k_{,l} = 0$. Differentiating the Jang equation, and using these
conditions yields $R^2k_{,ll} + 2R[R_{,l} + RK_R]_{,l} = 0$. Since the surface is the
outermost horizon we get $[R_{,l} + RK_R]_{,l} > 0$ so $k_{,ll} < 0$. Thus the
function decreases below the critical value of $k = 1$
as one moves out  and since there are no further
horizons we get a regular solution which asymptotes to zero. 

Now make the Jang transformation. The only metric component that changes is
$$
{\bar g}_{rr} = g_{rr} + f'^2 = g_{rr}(1 + f_{,l}^2) = {g_{rr} \over 1 - k^2}.
$$
Since $k = 1$ and $k_{,l} = 0$ at the horizon we have that $1 - k^2 \approx Al^2$
near the horizon for some constant $A$. This means that the proper distance in the
transformed metric from points `near' the horizon to the horizon itself becomes
infinite. However, the area of the spheres does not change because the angular
metric components are unaffected. This means that the manifold `near' the horizon gets
transformed into an infinitely long cylinder whose cross-section asymptotes to the
original area of the horizon and the three-scalar-curvature along the cylinder is a
positive constant ($^{(3)}\bar R \sim 1/R^2_H$; $R_H$ is the areal radius of the apparent
horizon). Now, however, when solving the
Lichnerowicz equation on this manifold we find that the conformal
factor cannot go to 1 at both ends since it behaves like $\exp (-Cl)$ near the horizon.
Such behaviour was first observed by Schoen and Yau in
\cite{SY2}. This means that, while a manifold with zero scalar
curvature can be constructed, we have no `inner' minimal surface whose area
approximates the original area of the horizon.

The second choice is to pick $k = -1$ at the horizon. If it is a future horizon we have
$k_{,l} > 0$ and there is a regular solution to the Jang equation. In addition, near
the horizon $1 - k^2 \approx Bl$. Now the radial metric, after the Jang
transformation, blows up at the horizon but does so in such a way that the proper
distance remains finite. Further, the inner surface becomes a minimal surface with the
same original area because the mean curvature scales to zero. While the
area of the nearby surfaces is unchanged, the proper distance between them becomes
large.  Again, when solving the Lichnerowicz equation with $\phi = 1$ at both ends we
expect
$\phi_{,l}$ to be negative at the inner boundary so in this case the mean curvature 
of nearby surfaces will become negative. Indeed, the Jang-transformed scalar curvature satisfies a positivity property of the form $^{(3)}\bar R\ge 2A_iA^i+2\nabla_iA^i$ with
$(A^i)=\left( kk'+k\left( 1-k^2\right) K_l,0,0\right) $. This is sufficient
to show that the manifold can be conformally transformed to one with zero scalar
curvature, i. e., we can solve $8\bar \nabla^2\phi -^{(3)}\bar R\bar \phi =0$, $\phi >0$ with $\phi =1$ at both ends. A solution can be expected rather like in the case
$^{(3)}\bar R\ge 0$, i. e., with an interior minimum. This means that $\phi $ decreases
at the inner minimal surface. The mean curvature of this conformally
transformed  surface becomes negative and a new minimal surface appears somewhere
outside it. Its area can be expected to be different from the area of the original apparent horizon; it can be smaller than the original area, if $^{(3)}\bar R\ge 0$, but since
we do not actually know the sign of the scalar curvature, there is no simple way of establishing which area is   bigger.

Thus in both cases, $k={^+_-}1$, there is no simple possibility
to exert the needed control  over  the asymptotic mass and
the area of outermost minimal surfaces.
We do not exclude the possibility that the Jang equation
 can be useful in order to
establish the validity of the Penrose inequality, but that would
 require a significant alteration and extension of the procedure.

\section{The existence problem in various gauges}

 As was mentioned in the introduction, one use of the Jang equation was in the Schoen
and Yau article \cite{SY3}. It is difficult to find
initial data which
 satisfy all the conditions laid down in the Schoen and Yau paper. Essentially the
only configuration we have
found is where one considers a spherical piece of a flat Friedmann universe glued to
some asymptotically flat data. The blowup of
the solution can be observed in this
situation.

 Let us  consider a flat spherical region of radius $R_0$, where the extrinsic curvature
is  pure constant trace, i.e., $K_{ab} = {K_0 \over 3}\delta_{ab}$. We seek a solution of
Eq.\ (\ref{J}) with the standard boundary condition of $k = 0$ at $R = 0$. There is a
natural scaling in the problem so we can choose $K_0 = -3$. This will gives a positive
solution and blowup occurs if and only if $k \ge 1$.

 Hence the equation we deal with is
 $$
 (R^2k)_{,R} - R^2[3 - k^2] = 0.
 $$
An upper bound can be found by setting $k^2 = 0$ and, since we are only interested in
the region
 where $ k \le 1$, a lower bound is obtained by setting $k = 1$. Therefore
 ${2 \over 3}R \le k \le R$. This means that for some value of $R$ which lies in the
range (1, 3/2),
 $k$ will pass through 1. Numerically, this has been found to occur at $R = 1.29$
\cite{Karkowski}. We have no
 contradiction to the Schoen-Yau result because the first apparent horizon is at $R =
1$. This
 calculation tells us that for any flat Friedmann sphere with $|K_0| = 3$ and
whose radius
 exceeds $R = 1.29$, independent of how it is extended into the asymptotically flat
region, the Jang equation has no regular solution.

 If we consider Theorem 2 of the Schoen-Yau  black hole paper, \cite{SY3}, and apply
their criterion for the absence
 of solutions to the Jang equation to this model, we find that they demand that no
solution
 can exist if $R_0 \ge \sqrt{2}\pi$. Again there is no contradiction, but note that it
is more than a
 factor of 3 bigger than 1.29.

 Another situation where  the spherical Jang equation can be easily analysed is on the
Painlev\'e-Gullstrand \cite{GP} slice of the Schwarzschild solution. This is the
 Schwarzschild slice where the spatial metric is
 flat and the extrinsic curvature satisfies

 $$ K_l = \sqrt{m \over 2R^3}, \hskip 1cm K_R = -\sqrt{2m \over R^3}.
 $$

 The Jang equation now becomes
 $$
 (R^2k)_{,R} - \sqrt{mR \over 2}(3 + k^2)  = 0.
 $$

 The solution of this equation at large $R$ must go like

 $$
 k \approx \sqrt{{2m \over R}}.
 $$

 From  Eq.\ (\ref{k}) it follows that 

 $$
 f_{,R} = {k \over 1 - k^2} \approx  \sqrt{{2m \over R}}.
 $$

 This can be integrated to give
 $$
 f \approx 2\sqrt{2mR}.
 $$

 We are interested in solutions that go to zero (or some constant value)
 at infinity. This solution does not satisfy this. Any data which asymptote to this
slice will have the same nonexistence property.

 Schoen and Yau demand that ${\rm tr} K \approx 1/R^3$. This would immediately exclude 
the Painlev\'e - Gullstrand data. The $1/R^3$ condition seems unnecessarily strong, it
may well be that ${\rm tr}K
 \approx 1/R^{2 + \epsilon}$ would suffice. The one thing that is clear is that some
falloff condition is required.

 The Jang equation may not possess a regular solution if the data contain
an apparent horizon. It is easy to show that the converse cannot hold. Consider any
moment of time symmetry data with a minimal surface, i.e., something like the moment of
time symmetry slice of the Schwarzschild solution, glued to some smooth interior. Then
the source term in the Jang equation vanishes and $f \equiv 0$ is obviously a regular
solution.

 In the spherically symmetric case consider some compact
  distribution of matter
 which is instantaneously at rest.   Eq.\ (\ref{J})
 reduces to $ (R^2k)_{,l} = 0. $
 Therefore there is a family of solutions $k = D/R^2$, where $D$ is any constant. The
only
 solution which is regular at $R = 0$ is the $D = 0$ one. Hence the $f = 0$ regular
solution is unique in the case of spherically symmetric moment-of-time-symmetry
 data.

Both the existence and uniqueness results extend to spherical maximal slices. The Jang
equation in this case is
\begin{equation}
(R^2k)_{,l} - R^2K_lk^2 = 0. \label{J4}
\end{equation}
Obviously $k = 0$ (and thus $f$ = constant) is a solution. This result also holds in the
nonspherical case. The only inhomogeneous term in the second
order Jang equation is
tr$K$. Therefore $f$ = constant is a solution in the maximal case irrespective of
whether the data contain apparent horizons or not.

In the spherical maximal case we can give a simple proof that $k = 0$ is the only
solution. Again, this is a proof by contradiction. Let us assume that a nontrivial
solution to Eq.\ (\ref{J4}) exists. This solution must vanish at the origin and let us
assume that it is positive, at least initially. In terms of the proper distance
coordinates, let us  choose an $l_1$ which satisfies two conditions. We want that $R^2k$
at $l_1$ is  the maximum over the interval $[0, l_1]$ and also that
$$
\int_0^{l_1}|K_l|k dl < 1.
$$
Let us integrate Eq.\ (\ref{J4}) over the interval $[0, l_1]$ to get
$$R^2k|_{l_1} = \int R^2K_lk^2 dl \le {\rm max}|R^2k|\int |K_l| k dl.
$$
This is a contradiction because $R^2k|_{l_1} = {\rm max}|R^2k|$. Therefore $l_1$
cannot exist and so also a nontrivial $k$. The only assumptions needed are that $K_l$
is finite and that the proper distance coordinates are well behaved.

\section{{\rm tr}$K$ and (non)existence of solutions}

It was demonstrated in the previous section that the Jang equation can always be
solved for spherical maximal data. Therefore it is interesting to investigate the role
that tr$K$ plays in allowing/preventing the existence of regular solutions. An easy place
to analyse this is in the `transverse' gauge, where one assumes that $K_l
\equiv 0$. In this case Eq.\ (\ref{J}) reduces to
\begin{equation}
(R^2k)_{,l} + R^2{\rm tr}K = 0.
\end{equation}
This can be integrated out to some radius $R = R_1$ from the center to give
\begin{equation}
R_1^2k(R_1) = -\int R^2{\rm tr}K dl.
\end{equation}
This now gives
\begin{equation}
k(R_1) = {- \int {\rm tr}K dv \over A},
\end{equation}
where the numerator is the proper volume integral over the sphere inside $R = R_1$
and $A = 4\pi R_1^2$ is the area of the surface of the sphere. Hence, if
$|\int {\rm tr}K dv| \ge A$ for any sphere, then we will not have a regular solution
while if $|\int {\rm tr}K dv| < A$ for all spheres we have no blowup.

The Hamiltonian constraint of general relativity can be used to show that this condition,
i.e.,
\begin{equation}
{|\int {\rm tr}K dv|\over A} \ge 1 \label{in}
\end{equation}
can only hold if a significant amount of matter is contained within the sphere.
The Hamiltonian constraint is
\begin{equation}
^{(3)}R - K^{ij}K_{ij} + {\rm tr} K^2 = 16\pi \rho.
\end{equation}
In the gauge where $K_l = 0$ this reduces to
\begin{equation}
^{(3)}R  + {1 \over 2}{\rm tr} K^2 = 16\pi \rho.
\end{equation}
If the scalar curvature is nonnegative this gives ${\rm tr} K^2 \le 32\pi \rho$.

Returning to the condition (\ref{in}), we can use the Schwarz inequality
 to get
\begin{equation}
1 \le {|\int {\rm tr}K dv|\over A} \le {\sqrt{\int {\rm tr}K^2 dv}V^{1 \over 2}
\over A} \le {\sqrt{\int 32\pi \rho dv}V^{1 \over 2} \over A}.
\end{equation}
Therefore, if the Jang equation does not have a regular solution, there must be a sphere
such that $M > A^2 /32\pi V$, where $M$ is the matter content.

In Sec. V we discussed the special case where the extrinsic curvature was pure
constant trace and found that if the region was large enough, regular solutions do not
exist. This result can be generalized to the situation where the extrinsic curvature is
pure trace, but not necessarily constant, i.e., $K_l = K_R$. Eq.\ (\ref{J}) becomes
\begin{equation}
(R^2k)_{,l} + R^2{\rm tr}K[1 - {k^2 \over 3}] = 0.
\end{equation}
Let us integrate this out to some radius $R_1$ and get
\begin{equation}
k(R_1) = -{ \int {\rm tr}K[1 - {k^2 \over 3}] dv \over A},
\end{equation}
Let us assume there exists a regular solution, i.e., $|k| < 1$ and that
${\rm tr}K$ has a fixed sign, say positive. One then gets
\begin{equation}
1 > { \int {\rm tr}K[1 - {k^2 \over 3}] dv \over A} > { 2\int {\rm tr}K
dv \over 3A}.
\end{equation}
Therefore, in a region which satisfies
\begin{equation}
{\int {\rm tr}K dv\over A} \ge {3 \over 2}, \label{in1}
\end{equation}
a regular solution to the Jang equation cannot exist. In the flat Friedmann model
we analysed in Sec. V we showed that a lower bound for this constant was 1.29.
This 3/2 may well be sharp. As above, this inequality can be related to the
requirement that the sphere contains a large amount of matter. A generalization of this
approach to nonspherical cases can be found in \cite{E}.

\section{Concluding remarks}

The Penrose inequality, $ A \le 16\pi m^2$, where $A$ is
the area of the outermost future
apparent horizon (with a non-negative optical scalar  $R_{,l} -RK_R$
 outward from the horizon) and $m$ is the ADM mass,
  is a consequence of cosmic censorship. Much
evidence for its correctness exists:
 it is true for spherically symmetric data
\cite{MOM94}; it is true in the moment-of-time-symmetry
case \cite{HI}; no numerical
counterexample exists in the general case. We believe it to be valid in general. However,
the Jang-equation-based scenario can not be used to prove it, since  it does not work 
in the spherically symmetric case (where   the Penrose inequality is correct).
A new idea is needed.

Another use of the Jang equation is in attempts to give a precise statement of the 
{\it trapped
surface conjecture} \cite{Seifert}  -- that the compression of matter leads
to the formation of an apparent horizon, as in the Schoen-Yau trapped surface article
\cite{SY3}. This is also problematic. 
   This equation is   -- as 
laid out in preceding sections --   rather insensitive to the existence of apparent
horizons.  To begin with, it does not  distinguish between future and past marginal
surfaces. This means that in order to obtain statements about the existence of apparent
horizons, one needs to control the sign of the quantity $R_l-RK_R$.
 This is done in
a recent analysis of this problem \cite{Yau}. Moreover, when  gauge conditions are
imposed which set the trace of the extrinsic curvature small enough
-- like the maximal slicing condition -- then the Jang equation
does not ``see" any horizons.  Clearly the  diagnostic
power of the Jang equation with regard to trapped surfaces is limited.

Having said that, it is necessary to point out that a demonstration of the trapped
surface conjecture   
remains elusive.    It has been well established in the spherically symmetric case
(\cite{BMOM}, \cite{BMOM88} -- \cite{Guven}) in spacetimes sliced with the
use of ``reasonable" gauge conditions, like maximal, constant mean curvature, flat or
polar slicings. The results concerning nonspherical situations remain patchy
(\cite{M91} --  \cite{Koc}  and references therein) and apply mostly to moment of time
symmetry initial data.  The conjecture itself seems to be self-evident. It is therefore
suprising that it is so difficult to formulate a clear quantitative description of this
initial phase of gravitational collapse. Seen from this perspective, the  approach
based on the Jang equation (\cite{SY3}, \cite{Yau}) is valuable in working in 
the ``large $trK$" sector of initial data, where it yields a quantitative description  of
a valid and interesting physical problem.

\begin{acknowledgements} 
N\'OM wishes to thank Rick Schoen for several enlightening conversations. This work was
partly supported by  the KBN grant 2 PO3B 00623 to EM. 
\end{acknowledgements}

\end{document}